\documentclass[12pt,preprint]{aastex}
\newcommand{\Msun}{~M_\odot}
\newcommand{\Rsun}{~R_\odot}
\newcommand{\msun}{M_\odot}

\newcommand{\lsun}{L_\odot}

\newcommand{\kms}{\rm ~km~s^{-1}}
\newcommand{\ergs}{\rm ~erg~s^{-1}}

\newcommand{\ml}{~\Msun ~\rm yr^{-1}}

\begin{document}

\title{SHOCK BREAKOUT EMISSION FROM A TYPE Ib/c SUPERNOVA: XRT 080109/SN 2008D}
\author{
Roger A. Chevalier\altaffilmark{1} and
Claes Fransson\altaffilmark{2}
}
\altaffiltext{1}{Department of Astronomy, University of Virginia, P.O. Box 400325, 
Charlottesville, VA 22904-4325; rac5x@virginia.edu}
\altaffiltext{2}{Department of Astronomy, Stockholm University, AlbaNova, 
SE--106~91 Stockholm, Sweden; claes@astro.su.se}

\begin{abstract}
The X-ray transient 080109, associated with SN 2008D, can be attributed to the
shock breakout emission from a normal Type Ib/c supernova.
If the observed emission is interpreted as thermal emission,
the temperature and radiated energy are close to expectations, considering that
scattering dominates absorption processes so that spectrum formation occurs
deep within the photosphere.
The X-ray emission observed at $\sim 10$ days is
attributed to inverse Compton scattering of photospheric photons with relativistic
electrons produced in the interaction of the supernova with the progenitor wind.
A simple model for the optical/ultraviolet emission from shock breakout is developed
and applied to SN 1987A, SN 1999ex, SN 2008D, and SN 2006aj,
all of which have optical emission observed at $t\sim1$ day.
The emission from the first three
can plausibly be attributed to shock breakout emission.
The photospheric temperature is most sensitive to the radius of the
progenitor star core
and the radii in these cases are in line with expectations from stellar evolution.
The early optical/ultraviolet observations of SN 2006aj cannot be accommodated by
a nonrelativistic shock breakout model in a straightforward way.
\end{abstract}


\section{INTRODUCTION}

The shock wave generated by the central explosion in a supernova accelerates
through the outer layers of the star, giving rise to the peak photospheric
temperature at the time of shock breakout.
The resulting X-ray/ultraviolet flash was predicted long ago \citep{KC78,Fal78,MM99}.
GRB 060218, discovered by its X-ray radiation \citep{Cam06}, was associated
with the fairly normal Type Ic SN 2006aj \citep{Maz06}, and its early emission
has been interpreted as due to shock breakout \citep{Cam06,Wax07}.
However, mildly relativistic motion is needed and it is likely that a
central engine played a role in the emission.

The X-ray transient XRT 080109 was serendipitously discovered during
{\it Swift} observations of SN 2007uy in NGC 2770 \citep{soder08}.
The supernova associated with the burst, SN 2008D, was 
found to be  
spectroscopically similar to the Type Ib SN 2005hg \citep{modjaz08}.
Early radio emission was detected, but not at the
high luminosity that would suggest relativistic motion, and
no gamma ray burst emission was detected by {\it Swift} \citep{soder08}.
In view of these properties, we further examine the possibility that the
early ($\la 1$ day) emission from SN 2008D was entirely due to the
supernova.
This view  was advocated by \cite{soder08}, but \cite{xu08} and \cite{li08}
find that a central engine flow is needed.
We treat the X-ray emission in \S~2 and the optical/ultraviolet emission in \S~3.
We take the distance to NGC 2770 to be 27 Mpc.

\section{EARLY X-RAY EMISSION}

The early X-ray emission from a supernova depends on the properties of the shock wave
generated by the central explosion.
The shock wave moving through the star is strongly radiation dominated,
giving rise to a burst of radiation when the shock wave reaches an
optical depth $\sim c/u$, where $c$ is the speed of light and $u$ is
the shock velocity.
The radiation accelerates the gas ahead of the shock wave, so that
the shock wave disappears; a gas dominated shock wave then forms 
in the surrounding medium when
the supernova radius has roughly doubled \citep{epstein81,fransson82}.
This evolution is shown well in the simulation by \cite{ensman94}, which
has parameters intended for SN 1987A.
The shock breakout begins at $t\approx 5.8\times 10^3$ s when the radiation
shock starts to spread, and reaches a peak luminosity at
$t\approx 1.0\times 10^4$ s when no shock is present.
The gas shock starts to develop at $t\approx 1.1\times 10^4$ s, when the
luminosity is down by a factor of 3 from its peak value, and
is well developed by $t\approx 1.7\times 10^4$ s, when the
luminosity is down by another factor of 3.
The simulation assumes pure absorption and an $r^{-7}$ density power law
for the preshock atmosphere,
but these properties do not appear to be crucial for the shock evolution.
Particle acceleration at the gas dominated shock gives rise to
the radio synchrotron radiation emission that is observed from Type Ib/c
supernovae, including SN 2008D.
\cite{CF06} found that for the conditions typically found in SNe Ib/c,
the X-ray emission from the later interaction is dominated by nonthermal mechanisms (inverse
Compton and synchrotron) as opposed to thermal emission.

{\it Swift} and {\it Chandra} observations of the XRT 080109 showed an early bright phase
lasting $\sim600$ s, followed by a slow, fainter decline \citep{soder08}.
We identify the early bright phase with the breakout of the radiation dominated
shock and the later emission with the interaction with the circumstellar
medium.
An important issue is the spectrum of the initial burst.
There is general agreement that a power law spectrum (with photon index $\Gamma=2.3$) provides 
a better fit than a blackbody (with $kT=0.73$ keV) \citep{xu08,soder08,li08,modjaz08}, although
a blackbody fit may still be acceptable \citep{xu08,modjaz08}.
\cite{soder08} find an absorption corrected fluence of $1.7\times 10^{46}$ ergs for 
the power law spectral fit.

Previous studies of shock breakout \citep{KC78,Fal78,MM99} assumed that the 
initial breakout radiation has a blackbody or dilute blackbody spectrum.
The fits to the observed spectrum of SN 2008D show a preference for a power
law spectrum, and
photon Fermi acceleration is invoked by \cite{soder08} to
explain the observations of SN 2008D, based on calculations by \cite{wang07}.
In this scenario, the photons gain energy by scattering across a 
semi-relativistic shock front.
The authors assume that an infinitesimally thin shock front is present,
apparently from the time that the radiation dominated shock front breaks
down at an optical depth of $c/u$.
As discussed above,
the breakout radiation is capable of accelerating the
matter ahead of the shock front so that the formation of the gas dominated
shock is delayed; the viscous shock starts to form only when the breakout
luminosity is dropping from its peak value and the optical depth at the shock is decreasing.

In view of the possible difficulty with producing a power law spectrum, we consider
the thermal interpretation of the spectrum.
\cite{MM99} estimated the breakout radiation expected from a relatively compact
star with a radiative envelope; they determine the temperature by setting the
postshock pressure $=\frac{1}{3}aT^4$, where $a$ is the radiation constant.
For the explosion parameters, we use the energy $E_{51}=3$, where $E_{51}$ is the
energy in units of $10^{51}$ ergs, and ejecta mass $M=4\Msun$, determined by
\cite{soder08} from the supernova light curve and spectra.
Taking the opacity, $\kappa$, of the He rich gas to be $0.2$ cm$^2$ g$^{-1}$, the radius
of the progenitor star required to produce $kT=0.73$ keV at shock breakout is
$R=1.5\times 10^{11}$ cm.
With these values of $E$, $M$, and $R$, the expected radiated energy at shock
breakout is $1.7\times 10^{45}$ ergs.
However, \cite{KC78} and \cite{ensman92} find that the observed color temperature
can be $2-3$ times the effective temperature if a blackbody were assumed because
the spectrum is formed at a relatively deep layer in the scattering atmosphere.
If the effective temperature is reduced by a factor of 2 to $0.36$ keV, the radius is
increased to $R=6.4\times 10^{11}{\rm~cm}=9\Rsun$, which corresponds to a breakout energy
of $1.9\times 10^{46}$ ergs.
\cite{Woo95} find that a He star with initial mass $4\Msun$ has a radius
of $9\Rsun$ at the time of carbon ignition; a He star can become extended
in its late burning phases.
With this radius, the peak shock velocity has $\beta\gamma\approx 0.6$, a
high, but nonrelativistic, velocity.

Although the thermal shock breakout model can reproduce the temperature and
radiated energy of the observed burst, it falls short on the timescale.
The expected timescale is $\approx R/c=5-20$ s, but the observed timescale is
$\sim 100$ s (taken as the time with flux above half the peak flux).
One possibility is that the photosphere is actually in a dense
stellar wind around the Wolf-Rayet star \citep{soder08}. 
Assuming a constant wind velocity the optical
depth to electron scattering in the wind is
$
\tau_e = 0.2 \mu_e \dot M_{-5} /( r_{12} v_{3}) 
$ where the mean molecular weight $\mu_e \approx 0.5$ for a
non-hydrogenic gas, $r_{\rm 12}$ is the radius in units of
$10^{12}$ cm, the wind mass loss rate $\dot M_{-5}$ in
$10^{-5}\ml$ and $v_{3}$ is the wind velocity in 
$10^3\kms$, normalized to a typical value for
Wolf-Rayet stars.  From the observed radio emission, \cite{CF06}
estimated  $\dot M_{-5}v_{3}^{-1}=2.8$ for SN 1994I
for an efficiency of magnetic field production in the shock front of
10\%; a lower efficiency would increase the wind density.  
A similar estimate for SN 2008D, using radio data from \cite{soder08},
gives $\dot M_{-5}v_{3}^{-1}=1.5$.
For these parameters,  we have $\tau_e=1$ at $r_{12}= 0.15$, which is smaller
than the estimated stellar radius, indicating that the freely
expanding wind is not optically thick.

However, the wind structure in the immediate vicinity of the star is
likely to be more complicated than a  $\rho \propto
r^{-2}$ wind. 
The radiative acceleration thought to be responsible for the
high velocity occurs over a region of several stellar radii
\citep[e.g.,][]{GH05,Li07},  leading to an
enhanced density. 
\cite{GH05} find that the acceleration up to $\sim 1000
\kms$ takes place over a factor of 2 in radius, while the final
velocity of $\sim 2000 \kms$ was only reached at more than 10 stellar
radii. 
Another important, but less understood, effect is connected to
pulsational instabilities. \cite{fadeyev03} find that for He
core masses below $10 ~ \msun$ and $L/M \ga 7 \times 10^3 \lsun/\msun$
the star is pulsationally unstable. Hydrodynamical calculations show
that these pulsations steepen to shock waves at the surface, which
levitates the outer regions. For a 10 $\Msun$ He core and a luminosity
of $2 \times 10^5 \lsun$ they find that the radius of the star expands
by a factor of two, and the scale height at the surface increases from
$6 \times 10^{-3} R_{\rm ph}$ to $\sim 0.23 R_{\rm ph}$. 
There is observational evidence for transient mass loss before some Type Ib/c and IIn supernovae, e.g., SN 2006jc \citep{Foley07}, on a scale of
$\sim 10^{16}$ cm.
In view of these issues and the uncertainty in the nonequilibrium radiation
field, the difference between the model and observed timescales may not
be significant.

A crucial aspect of our scenario is that the spectral formation and thermalization
occurs at a considerable depth in the supernova, so that the photons just 
scatter in the mass loss region.
The place of thermalization is characterized by
$\tau_{\rm scatt} \tau_{\rm abs} \approx 1,$
where $\tau_{\rm scatt}$ is the scattering optical depth and $\tau_{\rm abs}$
is the absorption optical depth.
We have 
$\tau_{\rm abs}/\tau_{\rm scatt} \approx \kappa_{\rm ff} / \kappa_{\rm T}$,
where the absorption is dominated by the free-free opacity $\kappa_{\rm ff} = 1.77 \times 10^{-2} 
\bar g n^2 T^{-1.5} \nu^{-2}$ in cgs units, $\kappa_{\rm T}$ is the Thompson opacity,
and $n$ is the density.
The probability for absorption becomes $
\tau_{\rm abs}/\tau_{\rm scatt} \approx 1.46 \times 10^{-23} \bar g n T_7^{-1.5} E_{\rm keV}^{-2}$ ,
where $T_7$ is the temperature in units of $10^7$ K, $E_{\rm keV}$ is the
photon energy in keV, and cgs units are used.
At 1 keV thermalization requires $n \ga 6.8 \times
10^{22}/\tau_{\rm scatt}^2$ cm$^{-3}$; photons will only scatter at lower densities
and thermalization  occurs in the steep density region.
We thus find that it is plausible that the temperature and energy can be reproduced,
although detailed hydrodynamic, radiative transfer calculations are needed to
show this definitively.
\cite{xu08} and \cite{li08} argued against a thermal model for the emission
based on the small radiating area derived from the luminosity and temperature.
Here we argue that the spectrum is diluted in the scattering atmosphere, so that
a larger radiating region can be accommodated.

In a model intended for SN 1994I, \cite{CF06} found that inverse Compton
emission dominates over days $1-20$, with an X-ray luminosity on day 1
of $\nu L_{\nu}\sim 1\times 10^{39}\ergs$.
The model luminosity evolves slowly because of the balance between the
shock wave expansion and the increasing optical luminosity, and on
day 10 is $\sim 4\times 10^{38}\ergs$.
The observed X-ray emission from SN 2008D by {\it Chandra} on day 10 
had a luminosity ($0.5-10$ keV) of $\sim 10^{39}\ergs$ \citep{soder08},
which corresponds to $\nu L_{\nu}\sim 3\times 10^{38}\ergs$
for a $L_{\nu}\propto \nu^{-1}$ spectrum.
There is reasonable agreement with the model expectations; the radio and
X-ray emission suggest wind conditions similar to those in SN 1994I on
a scale $\sim (2-3)\times 10^{15}$ cm.
\cite{soder08} also advocate an inverse Compton origin for this emission.
The \cite{CF06} model  predicts a drop in X-ray luminosity after day 10 when the
photospheric luminosity drops.
However, the supernova was close to the detection limit of {\it Chandra} on
day 10 \citep{soder08}, making it difficult to check this prediction.

\section{EARLY OPTICAL EMISSION}

As the supernova photosphere expands and cools, the emission at lower frequencies,
including optical, increases.
Optical emission related to shock breakout thus peaks at a later time, $\sim$ a day or days.
Such emission was seen in the early observations of SN 1987A \citep{Ham88}
and \cite{Che92} developed an analytical theory to describe
the emission.
The main approximation in this approach was that the supernova 
be well into its free expansion phase, which requires a relatively
small radius progenitor star.
The model should thus be applicable to Type Ib/c supernovae and we
consider the Type Ic's SN 1999ex \citep{Str02} and SN 2006aj \citep{Cam06,Mod06},
as well as SN 2008D.
In all these cases, there are observations at an age $\sim1$ day.
Here, the theory of \cite{Che92} is extended so it can be compared to
detailed observations.

After correcting for a typographical error,
equation (2.8) in \cite{Che92}
for the  density of outer freely expanding gas at velocity $v$ becomes
\begin{equation}
\rho=1.4\times 10^{98} E_{51}^{3.59} M_0^{-2.59} F_1^{7.18} t^{-3} v^{-10.18}{\rm~g~cm^{-3}},
\label{density}
\end{equation}
where $M_0$ is the ejecta mass in units of $\msun$  and $F_1$ is a factor
that describes by how much the velocity of an element of gas increases
after it has first been shocked.
This expression was derived for the case of SN 1987A, which had a progenitor
with a radiative envelope, like SNe Ib/c.
In a planar blow off $F_1 = 1.81$, but is reduced in
a spherical expansion.
Taking $F_1 =1.35$ leads to density and pressure profiles that agree with
results of \citet[][their equations 46 and 47]{MM99} for the outer parts of an exploded star with a 
radiative envelope, and we use that as a
reference value.

Given the density structure, the photospheric radius can be found by
integrating $\kappa\rho$ in to the point where $\tau=1$.
Taking $\kappa=0.2$ g cm$^{-2}$, 
the photospheric radius is then
\begin{equation}
r_{ph}=3.0\times10^{14}E_{51}^{0.39} M_0^{-0.28} (F_1/1.35)^{0.78}t_d^{0.78}{\rm~cm},
\label{rph}
\end{equation}
where $t_d$ is the age in days.
The velocity of freely expanding gas at the photosphere is
\begin{equation}
v_{ph}=3.4\times10^{4}E_{51}^{0.39} M_0^{-0.28} (F_1/1.35)^{0.78}t_d^{-0.22}\kms.
\label{vph}
\end{equation}
The luminosity at the photosphere can be found by allowing for a diffusion
wave to move back into the exploded star \citep{Che92}
\begin{equation}
L_c=3.3\times 10^{42} E_{51}^{0.91} M_0^{-0.74}R_{12} (F_1/1.35)^{-0.17}t_d^{-0.34}\ergs,
\label{lph}
\end{equation}
where $R_{12}$ is the radius of the progenitor star in units of $10^{12}$ cm.
Because the inner layers are shocked at a lower velocity and are observed at a
later time than the X-ray observations, the radiation field in this case can be approximated
by a blackbody. 
Using $L_c=4\pi r_{ph}^2 \sigma T_e^4$, where $\sigma$ is Stefan-Boltzmann's constant
and $T_e$ is the effective temperature of the photosphere, we have
from equations (\ref{rph}) and (\ref{lph})
\begin{equation}
T_e=1.4\times10^4 E_{51}^{0.03} M_0^{-0.04}R_{12}^{0.25} (F_1/1.35)^{-0.43}t_d^{-0.48}{\rm~K}.
\label{teff}
\end{equation}

These results can be compared with those of \cite{Wax07}, who also
assumed that the photosphere is in the outer shock accelerated part of
the supernova density profile, but otherwise used a different method.
The scaling with parameters is similar, although not exactly that
found here, and, for the same reference values,
\cite{Wax07} obtain $r_{ph}=3.2\times 10^{14}$ cm and $T_e=2.6\times 10^4$.
The agreement with the $r_{ph}$ found here is good, but the value of 
$T_e$ found here is smaller by a factor of 1.9, or the luminosity is smaller by 12.   
The method of \cite{Wax07} does not include radiative diffusion, whereas
that of \cite{Che92}, used here, does.
We find that the luminosity decreases by 1.9 because of diffusion,
for the standard parameters at $t_d=1$, a relatively small effect.

The photospheric radius and temperature determine the spectral luminosity
evolution, as shown in Figure 1 for 4  wavelengths relevant
to the {\it Swift} UVOT bands.
The results in Figure 1 assume the reference values for the
physical parameters ($E_{51}=1$, $M=1\Msun, R_{12}=1$), 
but can be used for other values by scaling
the luminosity by a factor $E_{51}^{0.88} 
M_0^{-0.68}R_{12}^{0.81}$ and the time by a factor
$E_{51}^{-0.06} M_0^{0.08}R_{12}^{-0.52}$.
When the wavelength is in the Rayleigh-Jeans part of the spectrum,
we have $L_{\nu}\propto t^{1.1}$.

For the theory described here to be applicable, the photosphere
must be within the steep power part of the supernova density profile.
If there is no surrounding medium, the limitation at the high velocity
end is the maximum velocity at shock breakout, which is
$v_{max}\approx 129,000 E_{51}^{0.58} M_0^{-0.42}R_{12}^{-0.32}\kms$ \citep{MM99}.
Using equation (\ref{vph}) and the reference values, this velocity corresponds
to the velocity at the photosphere at an age of 0.002 day.
If the supernova is surrounded by a dense, optically thick wind,
the time at which the theory becomes applicable is after
the shock wave has broken out from the wind.
Also, because the density distribution at small radii becomes flatter than the steep
power law, the actual
value of $r_{ph}$ should gradually become smaller than the value estimated
here.
To estimate this effect, we used the harmonic mean density profile of
\cite{MM99} (their eq. [46]) to estimate the value of $r_{ph}$ more accurately.
For the reference parameters
and an age of 1 day, the value of $r_{ph}$ is 20\% smaller with the
more accurate density distribution.

Another requirement of the applicability of the simple model is that the
opacity remain constant.
Recombination of He$^{++}$ to He$^+$ occurs at a temperature of a few $10^4$ K
at the densities of interest; this results in a decrease in the opacity by
a factor of 2.
A larger decrease in opacity occurs when the gas becomes neutral,
which occurs at $T\sim 1.6\times 10^4$ K.
The recombination adds internal energy to the gas, which can help
power the light curve.
These considerations show that the simple model for the early
light curve starts to break down at an age $\sim1$ day.

The earliest observations of SN 1987A by \cite{Ham88} are on day 1.14
when 
$L=4.9\times 10^{41}\ergs$, $r_{ph}=1.4\times 10^{14}$
cm, and $T_e=13,600$ K.
To apply the model developed here, we use the parameters
$\kappa=0.34$ cm$^2$ g$^{-1}$ (to account for a H and He composition), 
$E_{51}=1$, $M_0=16$, and $R_{12}=2.1$
with the result $L=5.5\times 10^{41}\ergs$, $r_{ph}=1.6\times 10^{14}$
cm, and $T_e=13,100$ K.
In this case, the parameters are fairly well determined by extensive
modeling \citep[e.g.,][]{imshennik89} and it appears that the model presented here can approximately
represent the early evolution.

There are good early limits on the brightness of the Type Ic SN 1999ex
because of the
occurrence of another supernova, SN 1999ee, in the galaxy,
implying that the age of the supernova was
$\la 1.05$ days when first observed.  
At that time,
$L=2.8\times 10^{41}\ergs$, $r_{ph}=1.2\times 10^{14}$
cm, and $T_e=12,900\pm 1400$ K \citep{Str02}.
\cite{Str02} compare the light curve of SN 1999ex to models of \cite{Woo87}
with $M=6.2\Msun$; with $E_{51}=1$, our model then  
gives $r_{ph} \approx 1.8\times 10^{14}t_d^{0.78}$ cm, which is in adequate
agreement with the observations.
The corresponding model value of $T_e$ is $13,000R_{12}^{0.25}t_d^{-0.48}$ K.

From {\it Swift} UVOT observations of SN 2008D, \cite{soder08} estimate
that $r_{ph} \approx 3\times 10^{14}$ cm and $T\approx 10^4$ K at an age
of $\sim 1$ day; the closest time of observation is 1.7 day, so these values must
be regarded as estimates. 
For $E_{51}=3$, $M=4\Msun$, and $R=9\Rsun$, our model gives
$r_{ph} = 3.1\times 10^{14}$ and $T=1.2\times 10^4$, in reasonable
agreement with the observed values.
We note that the  radius $R$ refers to the outer star layers where shock acceleration
and a steep density gradient occur.
There is the expectation in the model that the supernova should be cooling at this
age, which is consistent with the optical/ultraviolet observations \citep{soder08}.

For SN 2006aj, \cite{Cam06} estimate $r_{ph}\approx 3\times 10^{14}$ cm
and a temperature of 3 eV at $t=10^5{\rm~s}=1.16$ day, corresponding
to $T_e\approx 35,000$ K and $L\approx 1\times 10^{44}\ergs$,     
far more luminous than
either SN 1987A or SN 1999ex at a comparable age.
To compare the observations with models, we use 
$E_{51}=2$ and $M_0=2$ \citep{Maz06}.
The model reference properties on day 1.16 values are then $r_{ph}=3.6\times 10^{14}$
cm and $T_e=13,000$ K, corresponding to $L= 2.6\times 10^{42}\ergs$.
The value of $r_{ph}$ is in reasonable agreement
with observations, but
the model $T_e$ is low, leading to a significantly lower
luminosity than is observed.
The critical parameter for $T_e$ is the radius of the
progenitor star.
The required radius to produce the temperature is $R\approx 5\times 10^{13}$ cm,
which is the radius of an extended red supergiant star and much larger than that expected
for a Wolf-Rayet star.
As noted above, a helium star can become more extended in the evolution leading up to the
explosion if its initial mass is low \citep{Woo95}, but the
effect does not appear to be sufficient to explain the radius needed here.

In addition, the evolution leading up to   $10^5$ s can be considered.
The {\it Swift} UVOT light curves for GRB 060218 \citep[][see also Ghisellini et al.
2007]{Cam06} can be compared to the model curves in Fig. 1.
The cooling indicated by the evolution of the 1880 and 2510 \AA\ model
light curves matches that seen in the observations fairly well.
However, between $10^4$ and $10^5$ s, the observations show a turnover at
the longer wavelengths (3450 and 5440 \AA) that is not in the model curves
and cannot be attributed to the expected cooling of the photosphere.
The turnover indicates a decrease in the emitting area that is not expected
if the emission is from the outer part of the exploding star.
In addition, it can be seen that, although there is a roughly power law
increase in the fluxes at early times, the observed slope is shallower
than the $t^{1.1}$ dependence expected in the model. 
For these reasons, the problem of underproduction of flux in the model
at $10^5$ s is even more severe at early times \cite[see][]{Ghi07b}.
Although the early {\it Swift} UVOT observations of GRB 060218 show a
tantalizing similarity to expectations for early supernova emission, a shock breakout
model does not reproduce the observations in a straightforward way, indicating
that a central engine plays a role in the emission.

\acknowledgments
We are grateful to P. Chandra and A. Soderberg for discussions and information, and to
C. Matzner and the referee for comments.
This research was supported in part by 
NASA grant NNG06GJ33G, the Swedish Research Council, and Swedish National Space Board.


\clearpage

\begin{figure}[!hbtp]   
\epsscale{.80}
\plotone{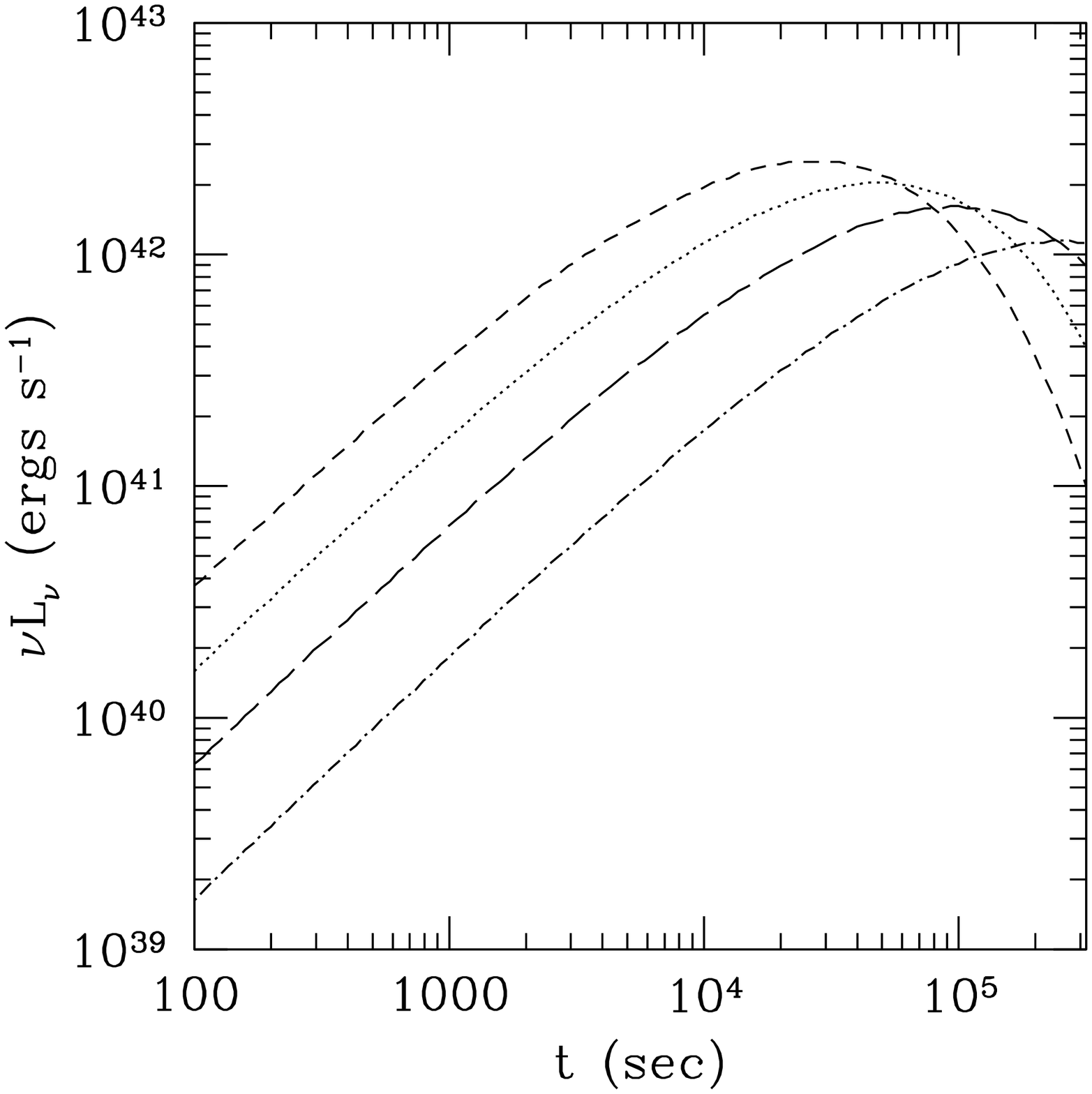}
\caption{Model predictions for the  photospheric spectral luminosity in the
shock breakout phase.
The curves are for 1880 \AA\ ({\it dash}), 2510 \AA\ ({\it dot}),
3450 \AA\ ({\it long-dash}), and 5440 \AA\ ({\it dash-dot}).
The model assumes reference values of the physical parameters, but
can be scaled to other values, as described in the text.
}
\end{figure}

\end{document}